\documentclass[
    ,final            
    , nomathfonts     
  ]
  {aipproc}

\usepackage{amssymb, amsmath} 

\layoutstyle{8x11single}

\begin{document}

\title{On the circular polarisation of light\\from axion-photon mixing}



\author{A.~Payez}{
  address={University of Li\`{e}ge, All\'{e}e du 6 Ao\^{u}t 17, 4000 Li\`{e}ge, Belgium}
}

\author{J.R.~Cudell}{
  address={University of Li\`{e}ge, All\'{e}e du 6 Ao\^{u}t 17, 4000 Li\`{e}ge, Belgium}
}

\author{D.~Hutsem\'ekers}{
  address={University of Li\`{e}ge, All\'{e}e du 6 Ao\^{u}t 17, 4000 Li\`{e}ge, Belgium}
}

\begin{abstract}
From the analysis of measurements of the linear polarisation of visible light coming from quasars, the existence of large-scale coherent orientations of quasar polarisation vectors in some regions of the sky has been reported. Here, we show that this can be explained by the mixing of the incoming photons with nearly massless pseudoscalar (axion-like) particles in extragalactic magnetic fields. We present a new treatment in terms of wave packets and discuss its implications for the circular polarisation.
\end{abstract}

\maketitle


\section{Introduction}

\begin{center}
			\begin{figure}[h]
				\includegraphics[trim = 10mm 10mm 0mm 0mm, clip, width=0.8\textwidth]{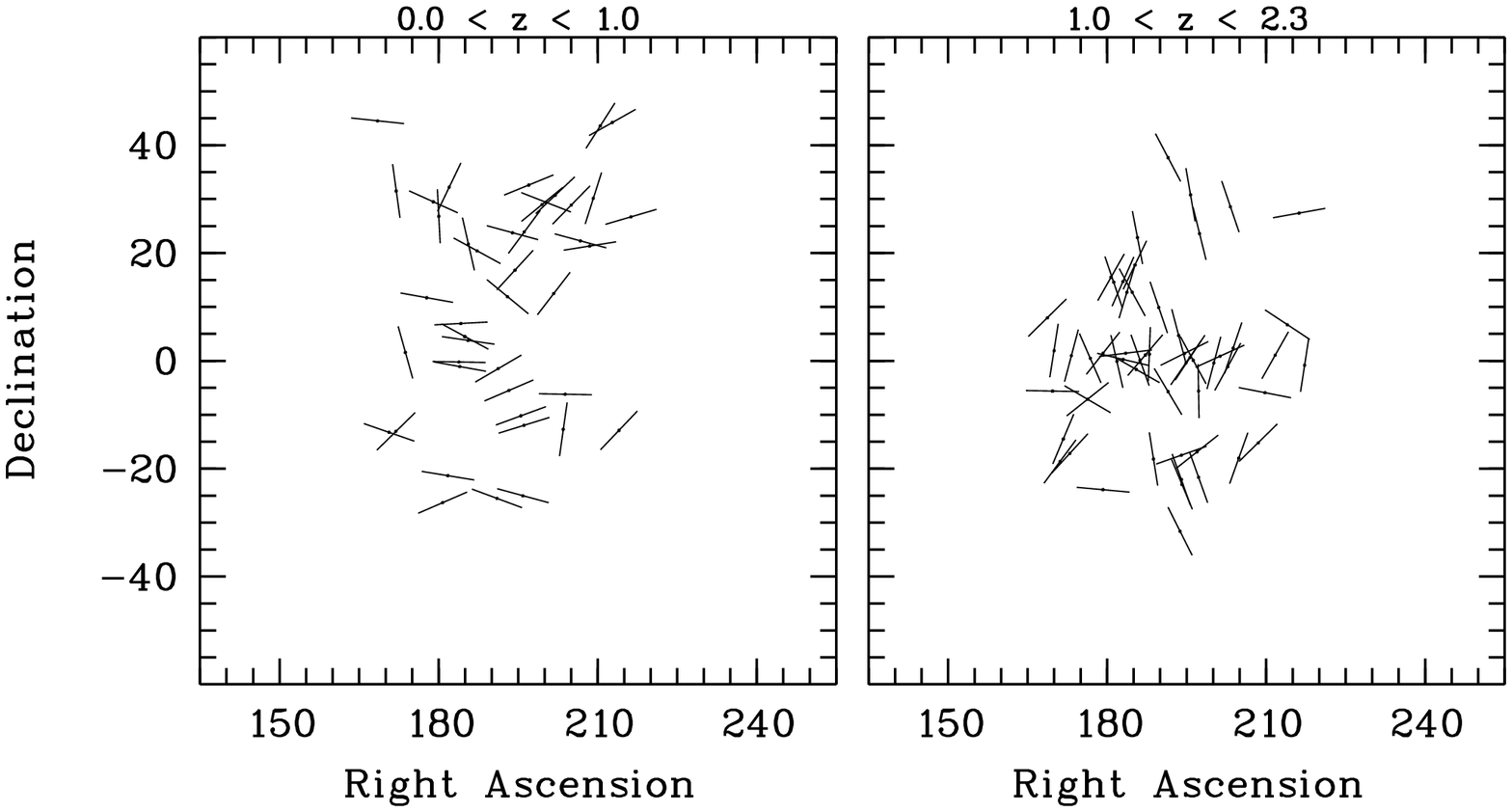}
				\caption{Maps of the same region of the sky in right ascension ($x$ axis) and declination ($y$ axis), both given in degrees, for AGN characterised by different redshifts, $z$. These polarisation vectors are thus for objects approximately along the same line of sight but at different distances from us. In the low-$z$ case, the average direction is $\bar{\theta}\approx 79^\circ$ while at high-$z$, it would be $\bar{\theta}\approx 8^\circ$ ($\theta$ being counted from North to East)~\cite{hutsemekers}.
				}
				\label{fig:hutsemekers}
			\end{figure}
		\end{center}

	In this work~\cite{paper}, we are looking at the effect that axion-photon mixing can have on the polarisation of light emitted by distant astronomical sources. In particular, the observation of redshift-dependent large-scale coherent orientations of AGN polarisation vectors~\cite{hutsemekers} can, at least qualitatively, be reproduced as a result of such a mixing of incoming photons with extremely light axion-like particles as they travel inside external magnetic fields. These observations (see Figure~\ref{fig:hutsemekers}) are based on good-quality measurements in visible light of the linear polarisation for a sample of 355 quasars from different groups, working on different instruments.

	This has been discussed in terms of axion-photon mixing by several authors, in the case of plane waves, and a prediction from this mixing, in this particular case and in general, is a circular polarisation comparable to the linear one and, hence, observable~\cite{planewaves,reviewHLPW}.
	Here, we present a treatment in which light is described by wave packets and show that the circular polarisation can be suppressed with respect to that which is predicted in the plane wave case.

\section{Axion-photon mixing: the basics}

	Although their existence is still hypothetical, from the theoretical point of view axions are well motivated \cite{w_w,P_Q} and many experiments ---using lasers~\cite{laser} in laboratories, or related to astrophysics~\cite{astro}--- are currently trying to detect them. When one relaxes the relation predicted between their couplings to other particles and their mass, and thus allows the consideration of the complete space of parameters, one then often talks about axion-like particles (ALPs). These ALPs are then, in general, not related to the resolution of the strong $CP$ problem but could appear for some other reason in extensions of the Standard Model. In the following, we discuss such particles, focusing ourselves on extremely light pseudoscalar axion-like particles.

	As already mentioned, the reason why axions and ALPs could explain the observation of large-scale coherent orientations of quasar polarisation vectors is their coupling to photons.
	This interaction is usually taken into account in an effective way by using the following Lagrangian density:

\begin{equation}
	\mathcal{L} =  \frac{1}{2}\ (\partial_{\mu}\phi) (\partial^{\mu}\phi) - \frac{1}{2}\ m^2\phi^2 - \frac{1}{4}\ F_{\mu\nu} F^{\mu\nu} - j_{\mu} A^{\mu} + \frac{1}{4}\ g \phi F_{\mu\nu}\widetilde{F}^{\mu\nu}, \label{eq:lagrangian density}
\end{equation}
	where $m$ is the mass of the axion field, $\phi$, and $g$ is its coupling constant with photons; $F^{\mu\nu}$ is the electromagnetic tensor and $\widetilde{F}^{\mu\nu}$, its dual. Note that the Lagrangian written in \eqref{eq:lagrangian density} is for a pseudoscalar axion-like particle; in the case of a scalar ALP, the coupling term $\frac{1}{4} g \phi F_{\mu\nu}\widetilde{F}^{\mu\nu} = - g \phi \vec{E}\cdot\vec{B}$ is to be replaced by $\frac{1}{4} g \phi F_{\mu\nu}F^{\mu\nu} = \frac{1}{2} g \phi (\vec{B}^2 - \vec{E}^2)$.

	The mixing of axion-like particles with photons is then usually discussed mathematically in terms of infinite plane waves. Using that description, the Stokes parameters can be computed; they are defined as:
                \begin{eqnarray}
                \left\{
                    \begin{array}{llll}
                        I  &=& E_{\parallel} E^*_{\parallel} + E_{\perp} E^*_{\perp}\\
                        Q &=& E_{\parallel} E^*_{\parallel} - E_{\perp} E^*_{\perp}\\
                        U &=& E^*_{\parallel} E_{\perp} + E_{\parallel} E^*_{\perp}\\
                        V &=& i(-E^*_{\parallel} E_{\perp}  + E_{\parallel} E^*_{\perp})
                    \end{array}
		\right.,
                    \label{eq:Stokes}
                \end{eqnarray}
where $I$ is the intensity and the other three are the unnormalised Stokes parameters: $Q$ and $U$ describe the linear polarisation and $V$, the circular one. In \eqref{eq:Stokes}, the indices parallel and perpendicular indicate the direction of polarisation of the electromagnetic radiation with respect to some arbitrary direction. In the case of axion-photon mixing, this direction is often chosen as that of the component of the external magnetic field transverse to the direction of propagation. With these parameters, predictions of the polarisation of light from the interaction in external magnetic fields can be given; the main properties of such a mixing being \emph{dichroism} and \emph{birefringence} (see~\cite{reviewHLPW} for a more detailed review of the plane wave case).

		\begin{center}
			\begin{figure}[h]
				\includegraphics[width=0.25\textwidth]{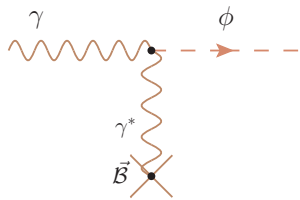}
				\caption{Axion-photon mixing.
				}
				\label{fig:primakoff}
			\end{figure}
		\end{center}

	Let us now recall that the selective absorption of one direction of polarisation which defines dichroism (see Figure~\ref{fig:primakoff}) will be different depending on the scalar or pseudoscalar character of the axion-like particle. Indeed, we have seen before that the coupling, in the pseudoscalar case, is proportional to $g\phi\vec{E}\cdot\vec{B}$, and the magnetic field can be decomposed as the sum of $\vec{\mathcal{B}}$, the external transverse magnetic field light is travelling through\footnote{Here, we neglect the influence of the longitudinal component of the electric field.} and $\vec{\mathcal{B}}_{r}$ ---$\vec{\mathcal{E}}_{r}$ and $\vec{\mathcal{B}}_{r}$ being the fields of the radiation. We immediately see that the coupling between such ALPs and photons is proportional to $\vec{\mathcal{E}}_{r}\cdot\vec{\mathcal{B}}$, namely that only photons polarised parallely to the external magnetic field, i.e. with $\vec{\mathcal{E}}_{r}$ parallel to $\vec{\mathcal{B}}$, will be affected by the existence of these pseudoscalar particles.
If the ALP is scalar, on the other hand, its coupling to photons will be proportional to $g \phi (\vec{B}^2 - \vec{E}^2)$. The $\vec{B}^2$ operator contains $\vec{\mathcal{B}}\cdot\vec{\mathcal{B}}_r$ so that, in order to have a non-zero mixing, we need $\vec{\mathcal{B}}_r$ parallel to $\vec{\mathcal{B}}$. Thus, as in an electromagnetic wave the electric field is perpendicular to the magnetic field, we understand why only photons polarised perpendicularly to $\vec{\mathcal{B}}$ will mix in that case.

		\begin{center}
			\begin{figure}
				\includegraphics[width=0.7\textwidth]{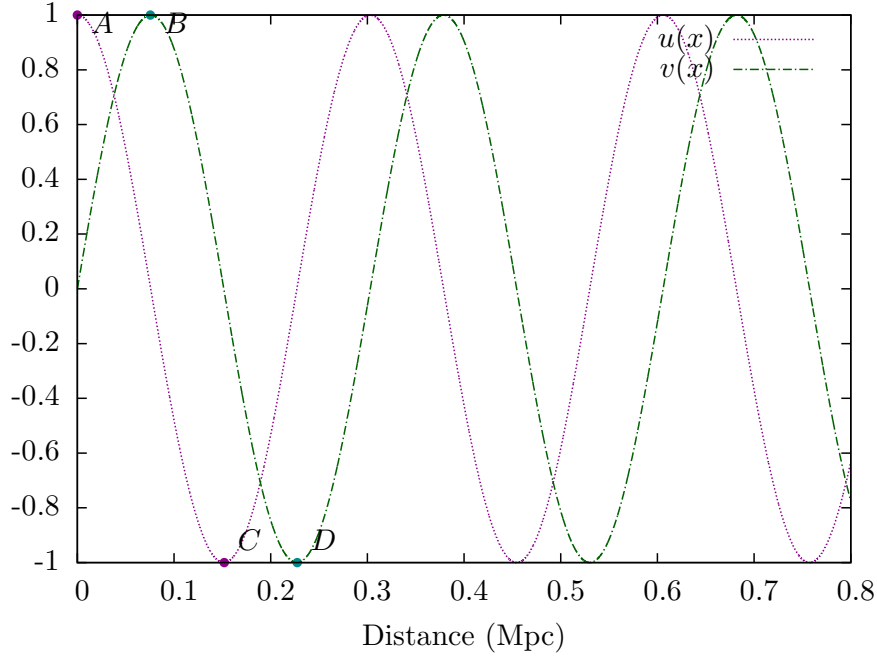}
				\caption{Evolution as a function of the distance traveled inside an external magnetic field, ${\mathcal{B}} = 0.1~\mu$G, of the circular polarisation, $v(x)$, and of one of the parameter describing linear polarisation, $u(x)$, in the case of a initially fully linearly polarised light ($\lambda = 500$~nm) described by a plane wave. Here, the plasma frequency, $\omega_p=3.7~10^{-14}$~eV, the mass of the ALP, $m=4.5~10^{-14}$~eV, and its coupling constant to photons is $g=7~10^{-12}$~GeV$^{-1}$.
				}
				\label{fig:uv}
			\end{figure}
		\end{center}

	This being, while dichroism would be an interesting way to produce linear polarisation and, in particular, to explain the observations concerning quasars, birefringence ---which is linked to the creation of circular polarisation--- would give a very clear signature of the mixing. An illustration of this birefringence is given in Figure~\ref{fig:uv}, which shows the evolution of the Stokes parameters $u=\frac{U}{I}$ and $v=\frac{V}{I}$ of a plane wave radiation initially fully linearly polarised ($u(0)=1$, $v=0$) as it travels inside an external magnetic field. What is shown in this figure is that ---because of axion-photon mixing---, as it propagates, the beam, being initially linearly polarised (point $A$, $u=1$, $v=0$), develops an ellipticity ($u\neq1$, $v\neq0$) up to the point where it is circularly polarised\footnote{At least if we do not take into account the evolution of the Stokes parameter $q$ which is linked to dichroism and of no importance when we discuss birefringence.} (point $B$, $u=0$, $v=1$). As the propagation continues, from circularly, it becomes once again linearly polarised but with a plane of polarisation perpendicular to the initial one (point $C$, $u=-1$, $v=0$), then elliptical again, then, circularly polarised (point $D$, $u=0$, $v=-1$), etc. This is completely equivalent to what one also obtains with a birefringent crystal.

This is related to the fact that, due to the interaction between photons and axions, the eigenstates of propagation have different masses and do not travel at the same velocity anymore, leading to a phase-shift between what we have called parallel and perpendicular polarisations of light. As the condition for linear polarisation is very strict ---the two perpendicular polarisations have to be oscillating exactly in phase---, the creation of ellipticity is a generic feature of the mixing mechanism.

An important result obtained in this formalism of plane waves is that, except in extremely specific cases\footnote{This is what one obtains if one does not assume very specific distributions of magnetic field orientations along the line of sight.}, the \emph{circular polarisation} predicted \emph{can be} \textit{a priori} \emph{as large as the linear polarisation}.

\section{Axion-photon mixing using Gaussian wave packets}

	The idea is to send wave packets into a region of uniform magnetic field and to compute the Stokes parameters. Before the magnetic field, the wave packets have the form:
	\begin{equation}
		E (x,t) = \int_{\omega_p}^{\infty} \frac{d\omega}{N}  e^{-\frac{a^2}{4}(\omega-\omega0)^2} e^{i \sqrt{\omega^2 - \omega_p^2} (x-x_0)} e^{- i \omega (t-t_0)},\label{eq:packet}
	\end{equation}
	where $\omega_p$ is the plasma frequency of the medium and $a$ controls the initial width of the packet (in the limit $a\rightarrow +\infty$, this reduces to the plane wave case).

	The main motivation for considering this formalism comes from the measurement of the circular polarisation of some of the quasars considered in~\cite{hutsemekers}. While axion-photon mixing would be an attractive explanation of the observations for linear polarisation, preliminary results indicate that circular polarisation of light from these AGN is, in general, \emph{much smaller} than the linear polarisation~\cite{polcirc}. This means that if the creation of circular polarisation was really a smoking gun of ALP-photon mixing, no matter how refined the description, these new observations would rule out the mixing mechanism and could only be used to constrain the existence of axion-like particles.

	For these reasons, it can be interesting to work with wave packets, as new effects will be taken into account, including dispersion, separation of packets and coherence; effects that might be of importance for the Stokes parameters.
	Note that calculating the propagation of packets of the form \eqref{eq:packet} is numerically\footnote{We use Multiple-Precision Floating-point library with correct Rounding: \url{www.mpfr.org}.} tricky, as the computation of the Stokes parameters requires a spatial resolution of the order of the width of the wave packets after a propagation over huge distances in the magnetic field (we will usually consider one magnetic field zone of 10~Mpc~\cite{vallee} and initial wave packets of width $\lesssim1\mu$m).

	\subsection{Results with wave packets}

		In the plane transverse to the direction of propagation, we choose a basis of two orthogonal linear polarisations, the same as the one used in the plane wave case, so that we will talk about polarisation parallel or perpendicular to the transverse external magnetic field $\vec{\mathcal{B}}$.
		This being done, we next choose the electric fields $E_{\parallel}(x,t)$ and $E_{\perp}(x,t)$ both initially described by a function of the form \eqref{eq:packet}. Then, we propagate these using the equations of motion
\begin{eqnarray}\left\{\begin{array}{llll}
(\square + \omega_p^2) E_{\perp}(x,t) &=& 0\\
(\square + \omega_p^2) E_{\parallel}(x,t) - g\mathcal{B} \partial^2_t\phi(x,t) &=& 0\\
(\square + m^2) \phi(x,t) + g\mathcal{B} E_{\parallel}(x,t) &=& 0
\end{array}\nonumber \right.,\end{eqnarray}
and find the expressions of the electric fields after a propagation, when axion-photon mixing is at work, inside a step-like magnetic field region.

		\begin{center}
			\begin{figure}
				\includegraphics[width=0.75\textwidth]{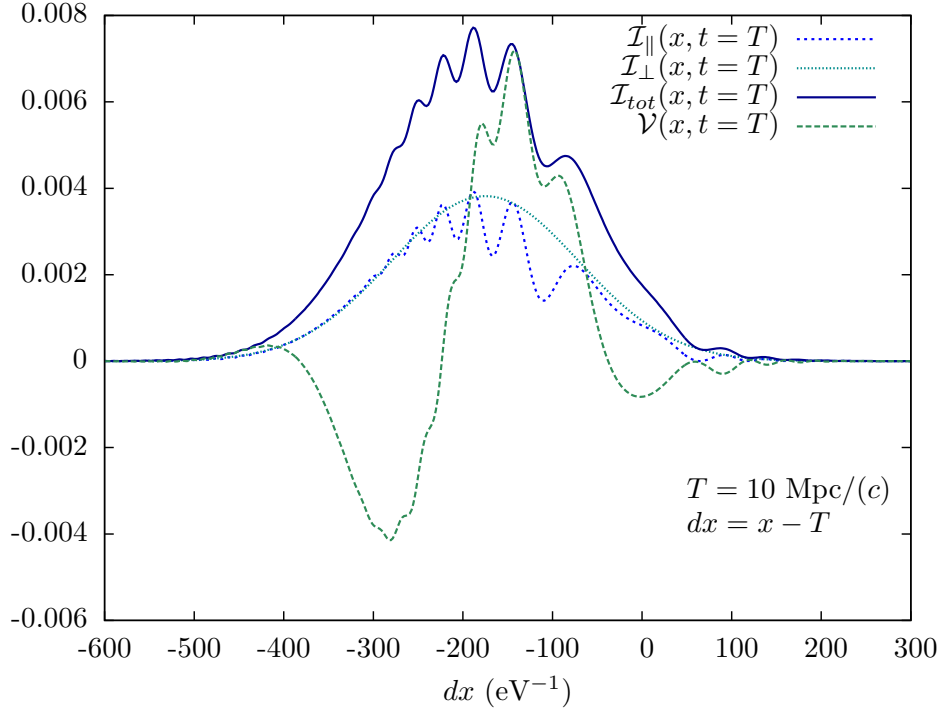}
				\caption{Intensities and unnormalised Stokes parameter $\mathcal{V}(x)$ after a propagation time $T$ in an external magnetic field, in a strong mixing case ---here, the axion mass is $m=4~10^{-14}$~eV, $\omega_p = 3.7~10^{-14}$~eV and $g\mathcal{B}=3~10^{-29}$~eV. The initial width of the wave packet has been chosen $\simeq\lambda_0$.}
				\label{fig:packets}
			\end{figure}
		\end{center}

		We can then use the expressions of the Stokes parameters~\eqref{eq:Stokes} ---which are observables built on intensities--- that can, for example, be plotted as functions of $x$, the distance travelled inside the magnetic field, for a given propagation time, $T$. This is what is represented in Figure~\ref{fig:packets} which shows what the two packets look like (respectively $\mathcal{I}_{\parallel}(x,t=T)$ and $\mathcal{I}_{\perp}(x,t=T)$) but also the total intensity (which is the sum of the two) and the unnormalised circular polarisation, $\mathcal{V}(x,t=T)$.
		This is for a beam with a central wavelength $\lambda_0 = 500$~nm, initially 100\% linearly polarised, with its polarisation plane making a 45$^\circ$ angle initially with the magnetic field direction (i.e. $u(0) = \frac{U(0)}{I(0)}= 1$; $q(0) = v(0) = 0$); this angle is, in fact, the most favourable one for the creation of circular polarisation, due to birefringence.
	Note also that the abscissa is $dx$, the position with respect to a frame moving at the speed of light $c$ (namely, a maximum at $dx = 0$ corresponds to $|\vec{v}|=c$).

		From the observational point of view, there is then a macroscopic exposure time over which one should integrate these functions to obtain, finally, the value of the observable Stokes parameters, e.g.:

		\begin{equation}
			V(x) = \int_\mathrm{exposure~time} dt~\mathcal{V}(x,t)\nonumber.
		\end{equation}

\begin{center}
			\begin{figure}
				\includegraphics[width=0.75\textwidth]{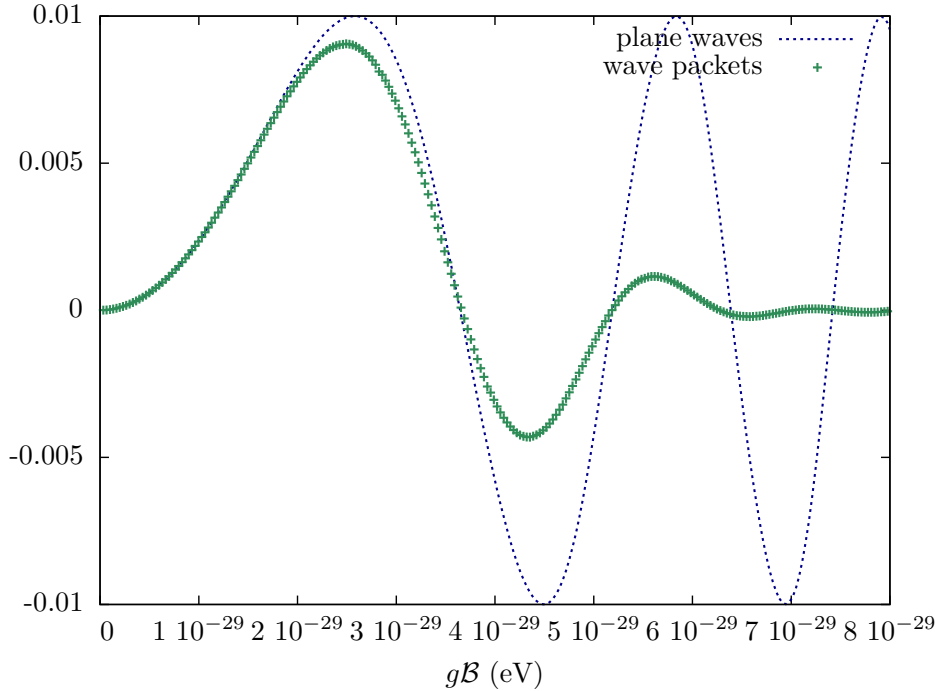}
				\caption{Circular polarisation for different values of the coupling in the case of initially partially linearly polarised light with $u(0)=0.01$. $\lambda_0$, $\omega_p$ ---and $a$, for the wave packet--- are the same as in Figure~\ref{fig:packets}, other parameters are: $T=10$~Mpc, $m=4.7~10^{-14}$~eV.
				}
				\label{fig:circpol}
			\end{figure}
		\end{center}

		From the computation of these integrals, we obtain that the wave packet formalism leads to a circular polarisation, $v=\frac{V}{I}$, \emph{lowered} with respect to the plane wave case. Figure~\ref{fig:circpol} illustrates the plane wave ($a\rightarrow\infty$) result: it shows the amount of circular polarisation generated by the axion-photon mixing with different values of the coupling $g\mathcal{B}$. In that simpler case, it is known that $v=\frac{V}{I}$ oscillates between $-|u(0)|$ and $|u(0)|$, whereas in the wave packet case it is shown that there is a damping of these oscillations. It follows from this observation that \emph{$v$ is no longer expected to be as large as the linear polarisation in general}.

\section{Conclusion}

	We have reviewed axion-photon mixing in the case of plane waves and have briefly presented our new formalism in terms of wave packets. The main consequence of this new treatment is the net decrease of circular polarisation with respect to what is predicted using plane waves. From this we conclude that the lack of circular polarisation in the light from AGN does not rule out the ALP-photon mixing.


\begin{theacknowledgments}
	A.~P. would like to thank the IISN for funding and to acknowledge constructive discussions on physical and numerical matters with Fredrik Sandin and Davide Mancusi. D.~H. is senior research associate FNRS.
\end{theacknowledgments}



\bibliographystyle{aipproc}   




\end{document}